\documentclass[11pt]{article}
\title{Some notable properties of the standard oncology phase I design}

\author{Gregory J. Hather \\ Department of Statistics, University of California, Berkeley, California, USA \\
\\ 
Howard Mackey \\ Genentech Inc., South San Francisco, California, USA \\ \\}

\date{\today}

\usepackage{fullpage}
\usepackage{graphicx}
\usepackage{amsmath}
\usepackage{bm}
\usepackage{setspace}

\begin{document}
\maketitle

\abstract{

We identify three properties of the standard oncology phase I trial design or 3 + 3 design. We show that the standard design 
implicitly uses isotonic regression to estimate a maximum tolerated dose. We next illustrate the relationship between the 
standard design and a Bayesian design proposed by Ji et al. (2007). A slight modification to this Bayesian design, under a 
particular model specification, would assign treatments in a manner identical to the standard design. We finally present 
calculations revealing the behavior of the standard design in a worst case scenario and compare its behavior with other 3 + 
3-like designs.

}

\section{Introduction}

\setlength{\parindent}{0.5cm}
\setlength{\parskip}{0.0cm}

The main goal of an oncology phase Ia clinical trial is to assess the safety of a drug which has not yet 
been tested in humans (Arbuck, 1996). A well designed oncology phase I trial should yield enough information
to determine a safe dose, or range of safe doses, to use in further trials, while maintaining a reasonably 
low level of risk to the patients in the study. The dose or doses to be used for further study should be 
low enough to be safe for most patients, but high enough to be efficacious, since higher doses are often more effective. 
While the `more is better' paradigm is well 
accepted in the case of chemotherapeutic agents, it is not clear that this paradigm should hold for newer, targeted, 
non-chemotherapeutic cancer agents. That is not to say that many or most of the newer molecularly targeted agents are not more 
efficacious in higher doses, but that it is not a given that the `more is better' paradigm always holds. 

The primary 
scientific question in an oncology phase Ia study dictates to some extent the type of patient who would enroll in such a 
study. Patients who have standard of care treatment available to them are less likely to participate in a phase I oncology 
study, and as a result, the patient population is often a heterogeneous group of patients with different types of late stage 
cancer. Different types of cancer often suggest different risk-benefit tolerances by both the patient and the treating 
clinician. This may result in a difficulty in selecting a trial design due to the existence and validity of multiple 
risk-benefit ratios. For example, a patient with metastatic pancreatic cancer may be willing to risk more toxicity than a patient 
with newly diagnosed late stage prostate cancer. This may result in the uncomfortable situation where the same adverse 
reaction has different implications for future development depending on the type of disease the patients has. In general, 
Phase I protocols report so called, dose-limiting toxicities (DLTs), irrespective of the type of disease the patient has. 
Additionally, the developers of a new oncology therapy may not know, before the first human data is generated in phase Ia, 
all the types of cancer to target in future phase II and III studies. This may be due to unexpected, 
dramatic results in phase Ia and/or changes in the financial resources of the entity developing the therapy. 
Another complicating reality is that assigning attribution of patient outcomes to the therapy under investigation, 
a patient's cancer, or a patient's concomitant medication is not an exact science. 
This is especially true when therapeutic agents are being tested for the first time in humans. 
Early on in testing, it is not uncommon for a patient's outcome to be deemed a DLT only after a number of other patients 
have experienced the same outcome and/or degree of severity. 
An outcome determined to be a DLT after patients have started treatment in higher dose levels may result 
in a protocol specified action which is different than the action already taken had the DLT been identifier earlier. 
These realities are rarely discussed in the literature for consideration of a 
phase I design, which further complicates the mission of designing the `best' phase Ia trial. 

Unlike the phase II or III 
paradigm, where the objective is to assess efficacy while obtaining valuable safety information, phase I studies often 
necessitate the administration of unsafe amounts of drug to some patients in order to determine a maximum tolerable dose, or 
range of safe doses. And as in all clinical studies, the trial should not involve too many patients or take too 
long to complete. 
One can find many methods for dose finding in the literature, the ``standard design'' (Storer, 1989) being the oldest of the 
commonly employed phase I designs. Other designs contained in the literature include continual reassessment (O'Quigley et 
al., 1990), random walk (Durham and Flournoy, 1994), escalation with overdose control (Babb et al., 1998), and cumulative 
cohort designs (Ivanova et al., 2007). For a more extensive review of various oncology phase I trial designs, the reader is 
referred to Rosenberger and Haines (2002), Potter (2006), and Koyfman et al. (2007). It should be noted that the standard 
design remains by far the most commonly used in the phase Ia oncology setting. This is likely due to the ease with which 
clinical centers are able to carry out the design, its simplicity, historical performance, and freedom from model 
assumptions. 

We have observed in the literature that comparisons made with newer competing designs, almost always using 
simulations, are usually performed with assumptions that do not reflect some common uses of the 3 + 3 design. One of these 
assumptions has to do with dose levels being fixed \textit{a priori} at the beginning of the study. For phase Ia studies 
especially, clinical investigators are usually not comfortable setting dose escalations before any human data are available. 
They prefer to fix the first dose level only and to have future dose escalations be determined by the accruing toxicity observed in the phase Ia. 
This is rightly due to the fact that pre-clinical toxicity information can perform poorly at 
predicting human toxicities, especially for newer targeted small molecule therapeutics. This enhancement to the 3 + 3 was 
introduced by Simon at al. (1997) as \textit{accelerated titration}, however its original incarnation referred to 
intrapatient escalations.  The authors' motivation for mentioning this fact is due to the belief that the 3 + 3, while 
possessing properties which would cause a formally trained statistician some discomfort (lack of an estimand to name one), it 
has served drug development relatively well. We make this comment in the context of new drugs that have yet to be tested in 
man, the phase Ia setting. We believe that progress in drug development methods will be surer as the professed merits and 
deficiencies of the 3 + 3 are more fairly assessed by accounting for more current clinical practice.

Recently, Ji et al. (2007) proposed a design similar to the cumulative cohort design, which uses a Bayesian model to describe 
the rate of toxicities at each dose. In this design, decisions about future doses are based on the posterior distributions of 
toxicity rates at current doses. These posterior distributions are a function only of the number of patients treated and the 
number of patients with toxicities at the current dose. Ji et al. noted that the decision rules for their design could be 
displayed in a monitoring table, where the columns correspond to the number of patients treated at the current dose level, 
and the rows correspond to the number of DLTs at the current dose level. Their work suggested new ways of thinking about 
phase I trial designs which inspired our investigation of the relationship between the standard design and their Bayes 
design. 
In the following sections, we will show that the standard design implicitly uses isotonic regression to estimate a maximum 
tolerated dose, is related to a special case of the Ji, Li, Bekele design, and possesses an analytical expression for the 
upper bound on the probability of selecting an unsafe final dose.  

\section{Background}

\subsection{Terms}

In the protocol of an oncology phase Ia study, a DLT will be defined as any of the 
pre-specified `dose limiting' adverse events that have been determined by the investigator to be related, 
or possibly related, to the drug and which occur within a specified temporal window. 
It should be noted that regulatory agencies are conservative with respect to assessing the safety of 
new therapeutic agents. As a result, any event occurring during the course of the DLT window that cannot be ruled out 
as being related to the agent is considered a DLT. These pre-specified `dose limiting' 
adverse events are defined as adverse reactions considered to be severe enough to limit a patient from 
further exposure to the experimental treatment. These definitions vary from study to study and are based upon 
the judgment of the authors of the protocol. A maximum tolerated dose (MTD) is the highest dose that is considered safe in patients. 
The standard design contains a heuristic, and the MTD is defined as the dose that is yielded by applying the heuristic. 
While the standard design yields an estimator for the MTD, an explicit estimand is not defined. Most other designs, 
including the Ji, Li, Bekele method, do 
define an estimand as well as its estimator for the MTD, the estimand being the dose 
at which a pre-specified fraction of patients experience a DLT in a well defined patient population. In this article, we use 
the term ``cohort'' to refer to 
a group of patients treated concurrently and ``dose group'' to refer to all the patients who receive a particular dose of 
drug. 

\subsection{The standard design}

The standard design, also commonly referred to as the 3+3 design, dictates either fully specified or 
partially specified dose level increases. A fully specified escalation scheme sets a starting dose level
and successively increases in future dose levels.  For example, dose level increases could either double the previous level 
or follow a modified Fibonacci sequence (Omura, 2003). A partially specified escalation scheme also specifies a starting dose 
level, 
however, future dose levels sequentially depend on the number and nature of toxicities experienced and/or 
pharmacokinetic or pharmacodynamic measures. In this article, we consider the fully specified escalation scheme 
with a pre-specified number of dose levels. 

The standard design is defined as having cohort sizes of 3 patients, where no dose level has more than 2 cohorts. 
This design begins treating patients at the lowest dose level. 
If none of the 3 patients in the first cohort experience a DLT, the next cohort of patients gets treated at the next higher dose 
level. If 1 of the 3 patients experiences a 1 DLT, the design assigns another cohort at the same dose level.  If, for some dose 
level,
2 or more (of 3 or 6) patients experience DLTs, then that dose level is considered unsafe and is not revisited again during the study, 
and the next cohort must be treated at the next lower dose.  
The trial ends when at most 
1 patient in 6 experiences a DLT, and 2 or more (of 3 or 6) patients experience DLTs at the next highest dose. 
The dose group that contains patients treated at the highest dose where at most 1 of 6 patients experience a 
DLT is determined to be the MTD. 
There is not a standard approach for addressing the scenario where the lowest dose level observes a DLT rate of more than 2 
out of 6, but two common approaches are to treat the next cohort at some reasonable fraction of the starting dose or to 
simply close down the study. 
Table \ref{standard-design} contains these rules of the standard design based upon the number of DLTs. 

While this design enjoys much popularity among investigators studying new cancer therapies, it does suffer from the 
previously mentioned defect that a well defined target, the estimand, is not known. In general, clinical investigators believe
that the standard design determines the dose which would cause between 17\% (1/6) and 33\% (2/6) of patients to 
experience a DLT. This belief is not completely misguided if one only considers the information contained 
in patients treated at the determined MTD to be relevant. However, a more sophisticated view would consider 
information contained at all tested dose levels. 
Under this perspective, the dose level that is estimated by the design 
is completely dependent on the true unknown DLT / dose relationship. For example, the standard design may yield an unbiased 
estimator of the dose which causes 20\% of patients to experience a DLT for one DLT / dose relationship versus 30\% under a 
different DLT / dose relationship. It has been reported in the literature that the standard design had targeted doses 
corresponding to a 23\% to 28\% DLT rate in 22 phase I studies (Smith et al., 1996). In addition, Lin and Shih (2001) 
reported that the standard design targeted DLT rates of 19 to 29\% based on 3 distinctly different dose / DLT relationships. 
Storer (2001) reports that the standard design is an appropriate design when interest is in identifying a dose which 
corresponds to a 20 to 25\% DLT rate.

It should be noted that while the standard design requires only 6 patients treated at a particular dose to determine the MTD, 
arguably too small a number to gain a reasonable understanding of an agent's toxic properties, common practice has become to 
treat an extra 12 to 18 patients at the MTD in order to gain more experience before designing a phase II clinical trial. An 
examination of how this practice affects the properties of this augmented standard design has not been examined in the 
literature to the authors' knowledge, and thus we will not discuss this case any further for the purposes of this article.

\subsection{The Ji, Li, Bekele method}

The Ji, Li, Bekele method uses a fully specified dose escalation scheme, and like the standard design, 
starts at the lowest dose and escalates dose levels by at most one dose level at a time.  
The decision 
for whether to escalate, stay the same, or de-escalate is based on the number of patients and the number of DLTs 
at the current dose.  A toxicity exclusion rule prevents dose escalation if the next higher dose is estimated to be 
unacceptably toxic based on patient outcomes at that dose.  
Cohorts of any size can be specified
and the stopping rule is based on a fixed pre-specified sample size,
with the one exception of early stopping when the starting dose is found unacceptably toxic.

Before we describe the method in detail, we would like to introduce some notation.  
First, we will use the integer $i$ to label the dose levels.  We define $p_i$ to be the 
true rate of DLTs at dose level $i$.  We let ${\bm p}$ denote a vector containing the values $p_i$ for all the dose 
levels.  Also, the Ji, Li, Bekele method requires a user-specified targeted ratio of DLTs which we shall denote by $p_T$. 

Using data that becomes available during the course of the study and a Bayesian framework with an assumed prior on ${\bm p}$, 
Ji, Li, and Bekele propose calculating a probability density for the probability $p_i$ seeing a DLT at the current dose $i$. 
The unit interval can be divided into 
$\{(0, p_T - K_1 \sigma_i), [p_T - K_1 \sigma_i, p_T + K_2 \sigma_i], (p_T + K_2 \sigma_i, 1)\}$, 
where $\sigma_i$ is the standard deviation of the posterior density of $p_i$, and $K_1$ and $K_2$ are parameters chosen by 
the researcher. 
If the first interval has the most posterior probability mass, then the dose is increased. If the second interval has 
the most posterior probability mass, then the dose is kept the same. If the third interval has the most 
posterior probability mass, then the dose is de-escalated. 

To better control the risk of treating patients at unsafe levels, Ji et al. included a toxicity exclusion rule, where one computes $P(p_i > p_T)$ 
conditional on the observed data.  If this probability is greater than some user defined quantity $\xi$, then the dose 
level $i$ and all higher doses are considered to be unacceptably toxic and those levels will not be visited again.  If 
the current dose is one level below an excluded dose, then the action is based on which of the first two intervals has the 
most posterior probability mass.  
The trial ends when the total number of patients used 
in the study reaches a pre-specified number or if the starting dose is found unacceptably toxic. 

At the end of the trial, the data and the prior are used to estimate the DLT rate $E(p_i) = \hat{p}_i$ at each dose level
not ruled out by the toxicity exclusion rule. 
Next, isotonic regression is performed on the vector of estimated DLT rates, ${\bm {\hat p}}$.
The isotonic regression procedure uses the weight $\sigma_i^2$ at each dose level $i$ and produces a new estimate ${\bm p^*}$ 
for the DLT rates that is a monotonically increasing sequence. 
The estimated MTD is the dose $i$ at which $|p^*_i - p_T|$ is minimized. 
If some doses tie for the smallest difference, and if the mean of $p^*_i$ among the ties is less than $p_T$, then
the highest dose among the ties is selected.  Otherwise, the lowest dose among the ties is selected.

\section{Isotonic regression and the standard design}

The isotonic regression estimator for the MTD has been proposed by Leung and Wang (2001), Stylianou and Flournoy (2002), 
Ivanova et al. (2003), and Ji et al. (2007). In simulations by Ivanova et al., an isotonic regression estimator  
outperformed both an empirical estimator and a parametric maximum likelihood estimator for the MTD.
We will now show that the standard design's method for estimating the MTD is equivalent to an estimator involving 
isotonic regression. 

In general, isotonic regression is a nonparametric regression technique that yields a fit to a vector which minimizes the 
residual sum of squares, subject to the constraint that the fitted values constitute a monotonically increasing (or decreasing) 
sequence (Robertson et al., 1998).  In other words, if ${\bm y}$ is a vector to be fit by isotonic regression, then the result would be a 
vector ${\bm z^*} = \mathrm{min}_{\bm z} ||{\bm y} - {\bm z}||^2$ subject to the constraint that $z_i \le z_j$ for all $i < j$.  One 
basic property of isotonic regression is that if there is some $i$ for which $y_j \ge y_i$ for all $j$ greater than $i$ and $y_j \le y_i$ 
for all $j$ less than $i$, then the fit at $i$ is $z^*_i = y_i$.

Isotonic regression is an appealing tool in the Phase I trials setting because the assumption of toxicity increasing with 
dose is usually quite reasonable.  After ending a Phase I trial, one can apply isotonic regression to the vector 
${\bm {\hat p}} = (t_1/n_1, t_2/n_2, \ldots, t_D/n_D)$ using the weights $(n_1, n_2, \ldots, n_D)$ where $t_i$ is the number 
of DLTs seen at 
dose level $i$, $n_i$ is the number of patients treated at dose level $i$, and $D$ is the total number of dose levels 
considered. 
These values and weights for isotonic regression have been used by other researchers to estimate the 
dose toxicity curve (Stylianou and Flournoy, 2002), in which  case, the isotonic regression estimator is equivalent to the 
order restricted nonparametric maximum likelihood estimator (Sun, 1998).  

After finding ${\bm {\hat p}}$, there are several ways to estimate the MTD.
Here, we consider the approach used by Stylianou and Flournoy (2002), where the 
MTD estimate is the largest dose with an estimated toxicity not exceeding a preset level.
Recall that when a trial using the standard design ends, the estimated MTD is the highest dose for which 6 people 
were treated and no more than 1 person experienced a DLT. We shall call this particular dose level $d$.
By the nature of the standard design rules, $t_i/n_i \leq 1/6$ for $i \leq d$ and $t_i/n_i \geq 1/3$ for $i > d$. It now 
follows that the isotonic regression estimate of the toxicity rate will be no more than $1/6$ for $i \leq d$ and at least 
$1/3$ for $i > d$.  Thus, if we set the target level $p_T$ to any value within the range $[1/6, 1/3)$, 
the dose level $d$ will be the largest dose with an estimated toxicity not exceeding $p_T$.   
Thus, for $1/6 \le p_T < 1/3$, this isotonic regression estimator of the MTD is equivalent to the MTD as defined by the 
standard design. 

\section{Dose assignment for the Ji, Li, Bekele  method and the standard design}

Ji, Li and Bekele note that when the priors used in their method are identical and independent among dose levels, then the 
action to be taken with respect to treating future patients depends on the cumulative number of patients and the cumulative 
number of DLTs in the current dose group. As such, the behavior of the design can be described by a trial monitoring table. 
In Ji et al. (2007), the trial monitoring table has columns corresponding to the number of patients treated at the current 
dose level and rows corresponding to the number of DLTs at the current dose level. These elements completely specify future 
action in all cases where the next higher dose level has not been found to have unacceptable toxicity.

With the standard design, future action also depends on the number of patients and the number of DLTs seen in the current 
dose group. Likewise, the standard design rules in Table \ref{standard-design} can also be described by a trial monitoring 
table, as in Table \ref{standard-design-alt}. We found that when the Ji, Li, Bekele method is used with cohorts of size 3, 
and parameters
$K_1 = 1$, $K_2 = 0.1$, $p_T = 0.17$, $\xi = 0.7$, and prior $B(0.005,0.005)$ for example, a monitoring table identical to 
Table \ref{standard-design-alt} was yielded. A range of parameter values close to the ones above also produced the same 
monitoring table. Thus, the actions during the trial are identical for both designs when the maximum sample size in the Ji, 
Li, Bekele method is replaced by a stopping rule of 1 DLT out of 6 patients with the next higher dose found to have 
unacceptable toxicity. That is, patients enrolled in either study would be assigned treatment in exactly the same way. In addition, 
while the MTD estimates of the two designs differ in general, they are identical for our choice of parameters.  This is 
because all the doses not excluded would have $\hat{p}_i$ less than or equal to 1/6.  Thus, all elements of ${\bm p^*}$ would 
be less than or equal to 1/6, which is less than $p_T = .17$.  Therefore, the highest non-excluded dose would be selected, 
which would be the same dose selected by the standard design.

We would like to suggest that a modification to the fixed stopping rule in the Ji, Li, Bekele method would not alter its spirit. 
Moreover, we expect that a more data driven stopping rule would be viewed as an enhancement by practitioners. 
As an example, if a dose was found to have unacceptable 
toxicity early under their design, it may dictate treating a large number of patients at the next lower dose level, 
regardless of how much information has been accumulated at that level.  
A data driven stopping rule might mitigate such an issue.

\section{The standard design and its potential to select an unsafe MTD}

There are many ways to assess how a clinical trial design would perform in terms of patient safety. Here, we consider the
standard design in terms of the underlying toxicity rate at the estimated MTD. While research exists with respect to this 
aspect in the form of potentially useful expressions (Reiner et al., 1999; Lin and Shih, 2001), our approach is to consider a 
worst case dose toxicity 
scenario. In particular, we consider a dose toxicity curve that maximizes the probability that the underlying toxicity rate 
at the estimated MTD
is at or above a given value.  
This consideration yields a simplified expression which can be used to, graphically or in tabular form, 
examine the behavior of the standard design, or modifications to it.  As a result, additional insight about the performance
of the standard design can be gleaned along with possible comparisons to other designs which are minor modifications 
to the standard design. These modifications could include changes in cohort size and/or the decision rule to expand 
or escalate/de-escalate, for example.  While some of these modifications are not used in practice, we introduce them only to 
illustrate that it is relatively simple to experiment with different design possibilities and examine their operating 
characteristics under this worst case.

Before proceeding, we introduce some notation.  Let $P(\textrm{DLT}_{\textrm{MTD}})$ be the underlying toxicity rate at the 
estimated MTD.  Let $r(v)$ be the probability that the standard design would pick an MTD for which $P(\textrm{DLT}_{\textrm
{MTD}})$ is greater than or equal to some value $v$.  Note that here, $v$ is not meant to convey a target rate of DLTs, as 
such a quantity is not defined for the standard design. Rather, $v$ is any DLT rate of general interest. For example, if a 
clinical investigator felt that a DLT rate of 30\% represented an unacceptably high DLT rate for a future cancer therapeutic 
in a particular cancer patient population, the investigator would be interested in the case where $v$ = 0.30. 

We wish to find an upper bound for $r(v)$.  
In order for $r(v)$ to be nonzero, the dose toxicity curve must have 
some dose level $d$ for which $p_i \geq v$ when $i \geq d$ and $p_i < v$ when $i < d$. To maximize $r(v)$, one must maximize 
the probability that 
at least one cohort is treated at dose level $d$.  One must also maximize the probability that the dose level does not 
de-escalate below $d$ given that at least one cohort is treated at dose level $d$. Thus, $r(v)$ is maximized when 
$p_i = 0$ for $i < d$ and $p_i = v$ for $i \geq d$.  
Note that $r(v)$ is not a complementary cumulative distribution function because the dose toxicity curve that maximizes $r
(v)$
is different for each value of $v$.  

Using the rules of the standard design, Appendix \ref{safety-calculations} shows that
\begin{eqnarray*}
r(v) = 1 - \frac{ 3 v (1-v)^2 (1-(1-v)^3) + (3 v^2 (1 - v) + v^3)}{1 - (1-v)^3 (3 v^2 (1 - v) + v^3)}
\end{eqnarray*}
We can also develop analogous expressions for modifications to the standard design. Consider the following hybrid 1+2+3/3+3 
design which is a variation of the two-stage design described in Storer (2001). One patient is enrolled at the starting dose 
level, and if a DLT is not observed, then another patient is treated at the next highest dose level. This process continues until 
one patient experiences a DLT, after which 2 more patients are treated at that dose, thereafter the standard design is used 
for the rest of the trial. The cohort sizes would then be 2 or 3, so as to make the dose group size divisible by 3 as in the 
standard 3 + 3 design. One can also consider other modifications to the standard design where the number of DLTs that dictate 
the next dose level to investigate are the same as that of the 3 + 3 but cohorts of size 2 or 4 are enrolled instead of 
cohorts of size 3. Of course, such designs will on average select different MTDs with different underlying DLT rates and will 
likewise have differing worst case scenarios.

\begin{figure}[hhh]
\caption{\doublespacing
$max$ P(choose MTD at a dose
  where $P(\textrm{DLT}_{\textrm{MTD}}) \geq v$) vs. $v$.  The designs are 3+3 (solid), 2+2 (dashed), 4+4 (dotted), and 
1+2+3/3+3 (dotdashed).  }
\begin{center}
\includegraphics[scale = 0.8]{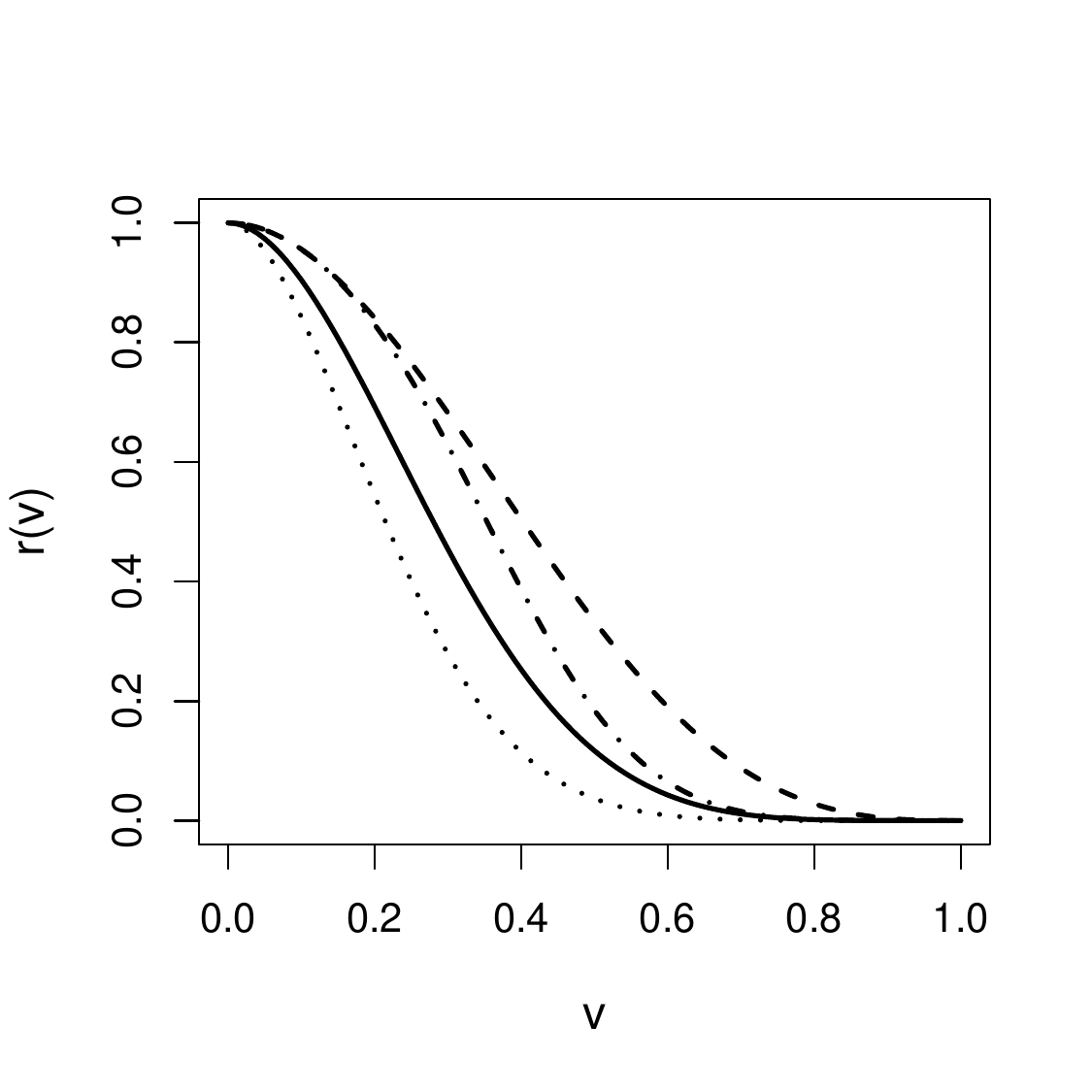}
\label{safety}
\end{center}
\end{figure}

In Figure \ref{safety}, we compare the behavior of these worst case scenarios which are the maximum probability of choosing 
an MTD with an underlying toxicity rate at or above $v$ as a function of $v$. Of the 4 designs described above, the 4+4 design 
appears to be the safest by this measure and 2+2 the least safe. Interestingly, the curve for the 1+2+3/3+3 design falls 
between the curve for the 2+2 design and the 3+3 design. It is also interesting to note that while the 2+2 is virtually 
indistinguishable from the 1+2+3/3+3 through probabilities up to 0.2, they become quite different at 0.6.

In addition to comparing the general behavior of different designs, these calculations may help guide in the choice of design 
for a specific phase I trial. For example, consider a trial involving a new cancer therapeutic that has shown promise in 
animal studies but which has the potential to cause severe DLTs based upon the mechanism of action of the drug. In this 
scenario, investigators would likely proceed cautiously with respect to how they treat the first patients with this drug and 
the manner in which they escalate dosing. In this case, oncologists may be particularly interested in an upper bound on the 
probability of selecting an unsafe MTD. If 0.25 is considered to be an unacceptably high rate of toxicity in 
this situation, we can see that the 3 + 3 design has, at most, a 57\% chance of selecting an unacceptably high MTD whereas 
the 1+2+3/3+3 has, at most, a 74\% chance. This observation, along with the designs' behavior in other scenarios, would 
presumably help inform the study design choice. Alternatively if we consider a cancer therapeutic with future plans to treat 
patients with late stage cancer where a more moderate rate of toxicity is acceptable, 0.35 may be considered to be the 
threshold for what is considered unacceptably high. 
If we consider the alternative concern of selecting a dose that is too low as the MTD, Figure \ref{safety} shows that the 4+4 
design always has at least a 30\% chance of choosing an MTD with a DLT rate below 0.15, even though higher doses are 
acceptable in these circumstances. This result may weigh against using such a design due to its relatively high risk of choosing an MTD that is too low, 
which could negatively impact efficacy.

\section{Discussion}

We have shown that the standard design implicitly performs isotonic regression to estimate the MTD.  In addition, the rules 
of the standard design can be described using a trial monitoring table of the form described by Ji et al. (2007).

Our results provide a more general way of viewing the standard design, which we hope will encourage 
further development and examination of oncology phase Ia designs while accounting for the realities of 
current oncologic practice. For example, one could change or add entries to the trial monitoring table (Table 2) 
or change the stopping rule to allow for more cohorts per dose level. One could also change the target rate 
for the isotonic regression or propose a completely new method for assigning treatment or determining an MTD. 
Drug developers and researchers are certainly ready and willing to consider new trial designs provided 
they truly improve existing methods.  

We have also illustrated the relationship between the 3 + 3 design and the Bayesian design introduced by Ji, Li, and Bekele, 
the relationship being that the 3 + 3 can be seen as a special case of the their Bayesian method with a modification to their 
stopping rule. 

In addition, we were able to analytically identify the upper bound for the probability of choosing an unsafe MTD 
with the 3 + 3 and similar designs. Similar analyses could straightforwardly be performed on other like designs like Storer's 
best-of-five design (Storer, 2001) for example. Examinations of these probability limits on more advanced designs, such as 
the Ji, Li, Bekele method, may provide useful insights, although it is not immediately obvious how to do so. 
The authors would like to also note that there are certainly more perspectives on how to assess the operating characteristics of a phase I trial 
design, as has been done in the literature. A common one being the expected number of patients treated at doses with high 
toxicity.  One could apply the framework presented here, considering the worst case scenario to find an upper bound on the expected 
number of patients treated at doses for which the probability of toxicity was at least $v$.

\section{Acknowledgments}

We are grateful to Mei Polley for referring us to the work of Ji, Li, and Bekele.  We also thank Wei Yu, Ron Yu, David 
Hiller, and Grazyna Lieberman for
their ideas and suggestions.  

\section*{References}

\setlength{\parindent}{0mm}
\setlength{\parskip}{0.3cm} 

Arbuck, S. G. (1996).   
Workshop on Phase I study design: Ninth NCI/EORTC New Drug Development Symposium, Amsterdam, March 12, 1996. 
{\it Ann. Oncol.} 7:567-573.  

Babb, J., Rogatko, A., Zacks, S. (1998). 
Cancer phase I clinical trials: Efficient dose escalation with overdose control.
{\it Stat. Med.} 17:1103-1120.

Durham, S. D., Flournoy, N. (1994).  
Random walks for quantile estimation. 
In: Gupta, S. S., Berger, J. O., eds. 
{\it Statistical Decision Theory and Related Topics V}.  
New York: Springer, pp. 467-476.

Ivanova, A., Flournoy, N., Chung, Y. (2007).  
Cumulative cohort design for dose-finding.
{\it J. Stat. Plan. Infer.} 137:2316-2327.  

Ivanova, A., Montazer-Haghighi, A., Mohanty, S. G., Durham, S. D. (2003). 
Improved up-and-down designs for phase I trials. 
{\it Stat. Med.} 22:69-82.

Ji, Y., Li, Y., Bekele, N. (2007). 
Dose-finding in phase I clinical trials based on toxicity probability intervals.  
{\it Clin. Trials} 4:235-244.  

Koyfman, S. A., Agrawal, M., Garrett-Mayer, E., Krohmal, B., Wolf, E., Emanuel, E. J., Gross, C. P. (2007).
Risks and benefits associated with novel phase 1 oncology trial designs. 
{\it Cancer} 110:1115-1124.

Leung, D. H. Y., Wang, Y. G. (2001).
Isotonic Designs for Phase I Trials.
{\it Controlled Clin. Trials} 22:126-138.

Lin, Y., Shih, W. J. (2001).  
Statistical properties of the traditional algorithm-based designs for phase I cancer clinical trials.  
{\it Biostatistics} 2:203-215.  

Omura, G. A. (2003).  
Modified Fibonacci Search.  
{\it J. Clin. Oncol.} 21:3177.  

O'Quigley, J., Pepe, M., Fisher, L. (1990).  
Continual reassessment method: a practical design for phase I clinical trials in cancer. 
{\it Biometrics} 46:33-48.

Potter, D. M. (2006).
Phase I Studies of Chemotherapeutic Agents in Cancer Patients: A Review of the Designs.
{\it J. Biopharm. Stat.} 16:579-604.

Reiner, E., Paoletti, X., O'Quigley, J. (1999).  
Operating characteristics of the standard phase I clinical trial design.  
{\it Comput. Stat. Data Anal.} 30:303-315.  

Robertson, T., Wright, F. T., Dykstra, R. L. (1988).
{\it Order Restricted Statistical Inference}.
Chichester: Wiley.

Rosenberger, W. F., Haines, L. M. (2002). 
Competing designs for phase I clinical trials: a review.
{\it Stat. Med.} 21:2757-2770. 

Smith, T. L., Lee, J. J., Kantarjian, H. M., Legha, S. S., Raber, M. N. (1996).
Design and results of phase I cancer clinical trials: three-year experience at MD Anderson Cancer Center.
{\it J. Clin. Oncol.} 14:287-295.

Simon, R. (1997).
Accelerated titration designs for phase I clinical trials in oncology.
{\it J. Natl. Cancer. Inst.} 89:1138-1147.

Storer, B. (1989).  
Design and analysis of Phase I clinical trials. 
{\it Biometrics} 45:925-937.  

Storer, B. (2001). 
An evaluation of phase I clinical trial designs in the continuous dose-response setting. 
{\it Stat. Med.} 20:2399-2408.

Stylianou, M., Flournoy, N. (2002).  
Dose finding using the biased coin up-and-down design and isotonic regression.
{\it Biometrics} 58:171-177.

Sun, J. (1998).  
Interval censoring.  
{\it Encyclopedia of Biostatistics}.  
New York: John Wiley \& Sons Ltd., pp. 2090-2095.

\setlength{\parindent}{0.5cm}
\setlength{\parskip}{0.0cm}

\appendix

\section{Worst case scenario calculations}
\label{safety-calculations} 

Consider the 3+3 design.  
Let $H$ be the index of the highest dose level visited during the trial, and let $K = H - d$.  Note that $H$ is unbounded 
from above, so $K$ is as well.  
Now,
\begin{eqnarray*}
r_{3+3}(v) &=& 1 - \sum_{k = 0}^{\infty} P(\textrm{0 DLTs for 3 patients})^k P(K = k | K \geq k) \\
        && P(\textrm{2 or more DLTs for 3 patients})^k \\
  &=& 1 - \sum_{k = 0}^{\infty} ((1-v)^3)^k \left[3 v (1-v)^2 (1-(1-v)^3) + (3 v^2 (1 - v) + v^3) \right] (3 v^2 (1 - v) + 
v^3)^k \\
  &=& 1 - \frac{ 3 v (1-v)^2 (1-(1-v)^3) + (3 v^2 (1 - v) + v^3)}{1 - (1-v)^3 (3 v^2 (1 - v) + v^3)}
\end{eqnarray*}
Likewise, for the 2+2 design,
\begin{eqnarray*}
r_{2+2}(v) &=& 1 - \sum_{k = 0}^{\infty} P(\textrm{0 DLTs for 2 patients})^k P(K = k | K \geq k) P(\textrm{2 DLTs for 2 
patients})^k \\
  &=& 1 - \sum_{k = 0}^{\infty} ((1-v)^2)^k \left[ 2 v (1-v) (1-(1-v)^2) + v^2 \right] (v^2)^k \\
  &=& 1 - \frac{2 v (1-v) (1-(1-v)^2) + v^2}{1 - (1-v)^2 v^2}
\end{eqnarray*}
and for the 4+4 design,
\begin{eqnarray*}
r_{4+4}(v) &=& 1 - \sum_{k = 0}^{\infty} P(\textrm{0 DLTs for 4 patients})^k P(K = k | K \geq k) \\
        && P(\textrm{2 or more DLTs for 4 patients})^k \\
  &=& 1 - \sum_{k = 0}^{\infty} ((1-v)^4)^k \left[ 4 v (1-v)^3 (1-(1-v)^4) + (1 - (1-v)^4 - 4 v (1-v)^3) \right] \\
  && (1 - (1-v)^4 - 4 v (1-v)^3)^k \\
  &=& 1 - \frac{4 v (1-v)^3 (1-(1-v)^4) + (1 - (1-v)^4 - 4 v (1-v)^3)}{1 - (1-v)^4 (1 - (1-v)^4 - 4 v (1-v)^3)}
\end{eqnarray*}
For the 1+2+3/3+3 design,
\begin{eqnarray*}
r_{1+2+3/3+3}(v) &=& 1 - \sum_{k = 0}^{\infty} P(\textrm{0 DLTs for 1 patient})^k P(K = k | K \geq k) \\
              && P(\textrm{2 or more DLTs for 5 patients})^k \\
  &=& 1 - \sum_{k = 0}^{\infty} (1-v)^k \left[ v (1 - (1-v)^5) \right] (1 - (1-v)^5 - 5 v (1-v)^4)^k \\
  &=& 1 - \frac{v (1 - (1-v)^5)}{1 - (1-v) (1 - (1-v)^5 - 5 v (1-v)^4)}
\end{eqnarray*}


\clearpage

\begin{table}[hhh]
\small
\caption{\doublespacing Rules for the standard design. The design stops enrolling patients after a dose is found where at 
most 1 of 6 patients experiences a DLT and where at least 2 DLTs are observed at the next highest dose. 
}
\begin{center}
\begin{tabular}{rcl}
DLTs & Patients	& Action \\
\hline
0 & 3 & Escalate \\
1 & 3 & No change \\
$\ge$ 2 & 3 & De-escalate \\
$\le$ 1 & 6 & Escalate \\
$\ge$ 2	& 6 & De-escalate \\
   \hline
\end{tabular}
\label{standard-design}
\end{center}
\end{table}

\clearpage

\begin{table}[hhh]
\small
\caption{\doublespacing Alternate depiction of the standard design rules.  The columns correspond to 
the number of patients treated at the current dose level, and the rows correspond to the number of DLTs at the current 
dose level. The elements specify the action, where `E' means escalate, `S' means stay the same, and `DU' means declare the 
current dose to be unacceptably toxic and de-escalate. The design stops enrolling patients after a dose is found where at 
most 1 of 6 patients experiences a DLT and where at least 2 DLTs are observed at the next highest dose. 
}
\begin{center}
\begin{tabular}{r|cl}
& 3 & 6 \\
\hline
0 & E & E \\
1 & S & E \\
2 & DU & DU \\
3 & DU & DU \\
4 &   & DU \\
\end{tabular}
\label{standard-design-alt}
\end{center}
\end{table}

\clearpage

\end{document}